\documentclass[preprint,showpacs,preprintnumbers,amsmath,amssymb]{revtex4-1}

\usepackage{graphicx}
\usepackage{dcolumn}
\usepackage{bm}
\usepackage[]{ams,amssymb,epsfig,color}
\newcommand{\be}{\begin{equation}}
\newcommand{\ee}{\end{equation}}
\begin{document}

\title{Chaotic spin-dependent  electron dynamics in a field-driven double dot potential}

\author{L. Chotorlishvili,$^{1,2}$ Z. Toklikishvili,$^{3}$ A. Komnik,$^{2}$ and J. Berakdar$^{1}$}
%

\affiliation{$^1$ Institut f\"ur Physik, Martin-Luther Universit\"at
Halle-Wittenberg, Heinrich-Damerow-Str.4, 06120 Halle, Germany\\
$^2$  Institut f\"ur Theoretische Physik, Universit\"at Heidelberg, Philosophenweg 19, D-69120 Heidelberg,Germany\\
$^3$ Physics Department of the Tbilisi State University, Chavchavadze
av.3, 0128, Tbilisi, Georgia}

\begin{abstract}
We study the nonlinear classical dynamics of an electron confined in a double  dot  potential and subjected
 to a spin-orbit coupling and a constant external magnetic
field.
It is shown that due to the spin orbit coupling, the
energy can be transferred from the spin to the orbital
motion.  This naturally heats up the orbital motion which, due to the
presence of the separatrix line in the phase space of the system,
results in a motion of the electron between the dots. It is shown
that depending on the strength of the spin orbit coupling and the energy of
the system, the electronic orbital motion  undergoes a transition
from the regular to the chaotic regime.
\end{abstract}
\pacs{} \maketitle
\section{Introduction}

The development  of  nanoscience, nanotechnology, and of the design of appropriate systems
for quantum information devices \cite{Raimond,Shevchenko,Zahringer,Wernsdorfer,Heinrich,Karabalin} have
triggered a large body of studies on the efficient and the controlled states preparation
 of systems at the nanoscale. In particular, magnetic nanostructures
were demonstrated to be highly promising both from an applied  \cite{Hirjibehedin,Rusponi,Mirkovic,Stroscio}
and fundamental nonlinear physics \cite{Chotorlishvili,Schwab} point of views.
In this respect  semiconductor quantum dots  with a spin-orbit coupling offer are a good case study \cite{Valin-Rodriguez,Levitov,Rashba}.
A key element in these systems is the coupling of the electron momentum to its
spin via  the spin orbit (SO) coupling. A momentum
dependent coupling offers a new way of manipulating the electron
spin by influencing the electron momentum via an external
periodic electric field. This is the idea of the electric-dipole
spin resonance proposed  by Rashba and Efros for the electrons
confined in lateral systems on the spatial scale between 10 and 100 nm \cite{Rashba}. However, a strong
external electric field can considerably influence the orbital dynamics and
drive the system far beyond of the linear regime \cite{Khomitsky}.
Usually nonlinearity implies a complicated   behavior and
difficulties may arise describing  the dynamics \cite{Khomitsky}. However, the nonlinearity
may lead, on the other hand,  to a variety of interesting  and subtle
phenomena  \cite{Reichl1984,Lin92,Sipe,Ngo} that have been discussed for a range of physical systems
 \cite{Raimond,Shevchenko,Zahringer,Wernsdorfer,Heinrich,Karabalin}.
 Two aspects are of interest here:
 First, what defines the regimes where
the systems exhibit strongly nonlinear behavior and possibly chaos, and
secondly, whether these nonlinearities and the chaotic behavior  can be utilized  for application
as   lead to   qualitatively new features in the  dynamics.

Here we address the first of these questions for the nonlinear electronic dynamics
where the driving force is a constant external magnetic field.
Thus, we consider  an  effect opposite to the electron-dipole spin resonance
protocol offered by Rashba and Efros \cite{Rashba}. In particular, we will consider the dynamics of the
electron with SO coupling term, confined in the quantum dot and
subjected to the action external magnetic field. Via  the
external magnetic field, one can directly control the spin dynamics
and therefore eventually influence the orbital motion.   We will show
that under different conditions, depending on the fields
parameters and the strength of the SO coupling, different types of
the dynamics can be realized.  These various types of the dynamics will
be linked to the structure of the phase space of the system. We
will demonstrate that using driving external fields one can switch the chaotic behavior  in the orbital motion
on and off.

\section{model}
We consider  a model system:  An  electron with mass $m_{e}$  is being confined to
a  double quantum well described by a  potential of the form
$U(x)=U_{0}\left[-2\left({x}/{d}\right)^{2}+\left({x}/{d}\right)^{4}\right]$.
Here $U_{0}$ is the energy barrier, and $2d$ is the distance between the minima.
We  will consider the low
temperature limit in which case the  orbital dynamics is not sensitive to the
thermally assisted tunneling. For the particular values of the
parameters $d\sim 100$  nm and $U_{0}\sim 20$ meV this imposes the
following restriction on the temperature $T<100$ K. For the
typical for GaAs electron effective mass
$m_{e}=0.067m_{0}$ the semiclassical tunneling probability is small, namely $\exp\left[-\frac{8\sqrt{2d\; m_{e}U_0}}{3}\right]\approx 10^{-4}$ .
Therefore, a classical consideration is justified. To quantify the SO
interaction we use the Dresselhaus type coupling term
$H_{so}=\alpha p\sigma_{x}$, where $p$ is the electron momentum, and we assume $\hbar\equiv 1$. Consequently the Hamiltonian of the system,
with  the applied external magnetic $B(t)$ being parallel to the $z$-axis, reads
\be
H=\frac{p ^{2}}{2m_{e}}+U(x)+\alpha p \sigma_{x}+
\mu_{B}g B(t) \frac{\sigma_{z}}{2}, \label{eq:ham}
\ee
where $\mu_{B}$ is the Bohr magneton, and $g$ is the electron Land\'{e} factor.
Introducing the characteristic maximum momentum
$p^{\max}=\sqrt{2m_{e}U_{0}}$ we can estimate the maximal
precession rate due to the SO coupling term
$\Omega_{so}^{\max}={2\alpha}\sqrt{2m_{e}U_{0}}$.
 While a constant magnetic field of the amplitude $B(t)=B$, induces a spin precession around $z$-axis with the frequency
 $\Omega_{B}={\mu_{B}g B}$.
 Therefore, if the amplitude of the magnetic field is strong enough
 $B>{2\alpha}\sqrt{2m_{e}U_{0}}/{\mu_{B}g}$,
 $\Omega_{B}>\Omega_{so}^{\max}$ the spin dynamics is described by the relation
 $\sigma_{x}(t)=\sigma_{x}^{(0)}\cos(\Omega_{B}t)$ and the Hamiltonian (\ref{eq:ham}) takes on the form
\begin{eqnarray}\label{eq:ham1}
&&H=\frac{p^2}{2m_{e}}+U(x)+\alpha p  \cos (\Omega_{B} t), \\
&&\sigma_{x}(t)\equiv\cos(\Omega_{B}t),\quad\sigma_{x}^{(0)}=1.\nonumber
 \end{eqnarray}
In what follows, we will the treat time dependent term in
(\ref{eq:ham1}) as a perturbation and for the sake of convenience we use
dimensionless units by introducing  the scaling: $E\rightarrow
E/4U_{0}$, $x\rightarrow x/d$, $t\rightarrow \Omega_{B} t$,
$p \rightarrow{p }/\sqrt{2mU_{0}}$, $\alpha\rightarrow\alpha\sqrt{{m_{e}}/{4U_{0}}}$.

\section{Solution of the autonomous system: Structure of the phase space}
\begin{figure}[t]
 \centering
  \includegraphics[width=8cm]{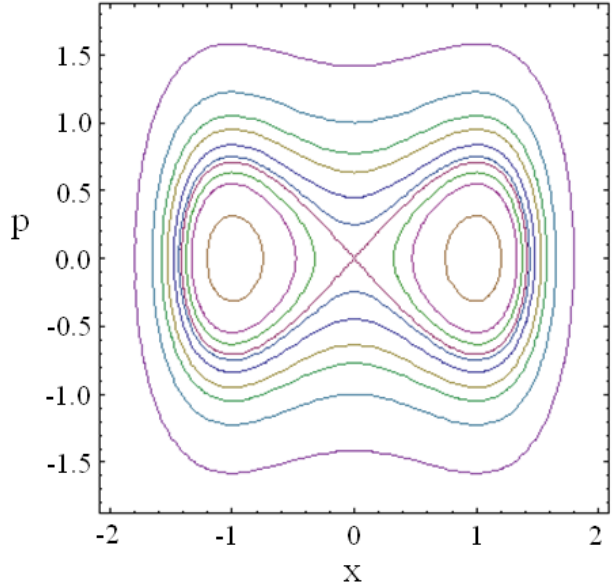}
  \caption{Phase portrait of the system (\ref{eq:motion}). The  phase trajectories with negative or positive values of the coordinate  $x<0$, $x>0$ describe the dynamics of the
  electron located on the left or the right well, respectively.
   These phase trajectories are separated by a separatrix line from the trajectories corresponding to the positive energy case. The separatrix crossing the nodal point $(p=0,x=0)$
   belongs to the area of a maximal chaos and arises if an external perturbation is applied.} \label{Fig:1}
\end{figure}

Before treating the general time dependent problem, we consider the
autonomous system and  find  the corresponding solutions. This allows to  identify the topological
structure of the phase trajectories. For the autonomous case $\alpha=0$ the
equation of motion corresponding to the Hamiltonian (\ref{eq:ham1}) reads
\be \dot{p}_{x}=-\frac{\partial H}{\partial x}=x-x^3 .
\label{eq:motion}
\ee
Performing the integration in
(\ref{eq:motion}) and inverting the result, for energies $E$ near the potential
minimum we obtain an oscillatory behavior in the form
\be
x(t)=\pm
\sqrt{x_{1}^{2}+(x_{0}^{2}-x_{1}^{2}){\rm cn}^{2}\left(x_{0}t/\sqrt{2};k\right)},\quad
-\frac{1}{4}<E<0 . \label{eq:solition}
\ee
Here $x_{0,1}$, $x_{0}=\sqrt{1+\sqrt{1+4E}}$, $x_{1}=\sqrt{1-\sqrt{1+4E}}$,
$x_{1}^{2}=2-x_{0}^{2}$ are the turning points defined by the
relation $V(x)=E$, ${\rm cn}(\ldots)$ is the elliptic Jacobi cosine
\cite{Abramowitz}, and the parameter $k$ has the form
$k^2=\left(x_{0}^{2}-x_{1}^{2}\right)/{x_{0}^{2}}$. Close to the
separatrix we have $E=0$  and consequently   $k=1$.  If the energy of the
system is negative, then the electron is located in the left or in the
right well and performs  oscillation confined by the potential barrier.
The oscillations in the different wells
are described by the solutions with different signs in
(\ref{eq:solition}). The time of the oscillation is given by
$T_{0}={2\sqrt{2}}K(k)/{x_{0}}$, where $K(k)$ is the complete
elliptic integral of the first kind \cite{Abramowitz}. If the
energy of the system is positive, the solution of the equation
(\ref{eq:motion}) reads
\be
x(t)=x_{0}{\rm cn}\left(\sqrt{x_{0}^{2}-1}t,1/k\right),\quad E>0 .\label{eq:solition1}
\ee
From the solutions (\ref{eq:solition}), (\ref{eq:solition1}) we
conclude that, depending on the values of the parameter $K$
the dynamics has qualitatively different nature. They are separated
by the value $K=1$ of the bifurcation parameter hinting
on the presence of  topologically distinct solutions (see
Fig.\ref{Fig:1}). In the case of a positive energy, the electron can
overcome the potential barrier and undergoes  inter-minima transitions, e.g.  from the  left
to the  right minima. The separatrix line that divides different
types of solutions corresponds to the zero energy case $E=0$ and the
separatrix solutions have the following form:
\be
x=\frac{\sqrt{2}}{\cosh(t)},\qquad p =-\frac{\tanh(t)}{\cosh(t)}.
\ee

\begin{figure}[t]
 \centering
  \includegraphics[width=8cm]{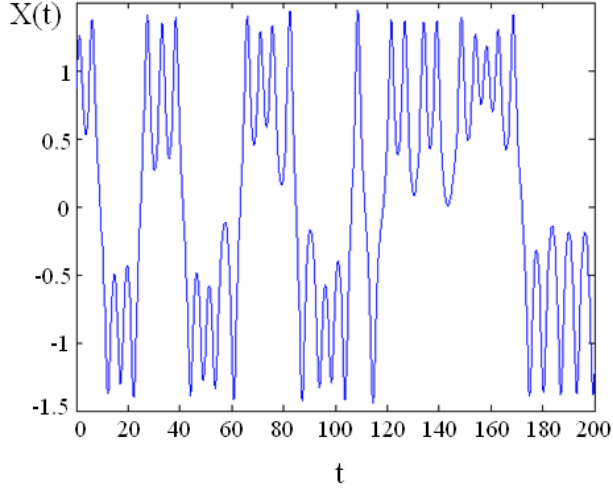}
  \caption{Numerical solutions of the system described by eq. (\ref{eq:ham1}). The generated chaotic trajectory lies close to the separatrix $E=0.01$, $\Omega_{B}=1$, $\alpha\rightarrow\alpha/4U_{0}=0.2$} \label{Fig:2}
\end{figure}

\begin{figure}[t]
 \centering
  \includegraphics[width=8cm]{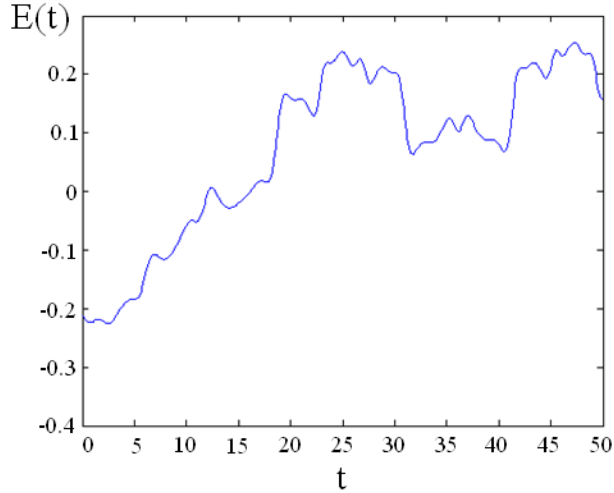}
  \caption{In order to reach the separatrix value of the energy $E\approx0$, the
  system needs a time which is clearly  larger than the inverse precession frequency  $t>1/\Omega_{B}$, $\Omega_{B}=1$, $\alpha=0.2$} \label{Fig:3}
\end{figure}

The dynamics near the separatrix is very sensitive  to
small perturbations. This fact can be exploited to help   the electron
overcomes the potential barrier and to perform a transition between two
equilibrium minima. The role of such a perturbation in our case is
played by the  SO term in (\ref{eq:ham1}).  The physical mechanism behind  the switching
of the electron position between the minima points is the formation of a
homoclinic structure. Homoclinic structure is formed due to the
applied perturbation in the vicinity of the separatrix crossing the
nodal point. The formation of the homoclinic structure can be verified
via the Melnikov function method \cite{Zaslavsky}. The equations of
motion for a perturbed system can be represented in the following
general form
\begin{eqnarray}
\dot{x}=\frac{\partial H_{0}}{\partial p }+\alpha f(x,p ,t) , \nonumber \\
\label{eq:pert} \dot{p}_{x}=-\frac{\partial H_{0}}{\partial
x}+\alpha g(x,p ,t) .
\end{eqnarray}
Here the terms  ${H_{x}=\partial H_{0}}/{\partial x}$,  ${H_{p }=\partial H_{0}}/{\partial p }$ describe the unperturbed motion, while
the contribution from the spin-orbit coupling are contained in  the terms
$\alpha f(x,p ,t)$ and $\alpha g(x,p ,t)$. Taking into account  (\ref{eq:ham1}) we immediately see $f=\cos(t)$ and $g=0$. Using
(\ref{eq:pert}) we can write Melnikov's  integral in the following
form
\be \label{eq:melnikov}
\Delta(\theta)=\int\limits_{-\infty}^{+\infty}
\left[H_{x}(x,p )f(x,p ,t)+H_{p }(x,p )g(x,p ,t)\right]dt .
\ee
Melnikov integral is a  measure of the distance between perturbed
stable and perturbed unstable separatrix trajectories.  Therefore, a
change of the sign of the Melnikov integral is equivalent to the
crossing of the stable and unstable separatrix trajectories.
Consequently, the formation of the homoclinic structure can be
identified as the change of the singe of the Melnikov integral.
On  the other hand, a crossing of the separatrix is equivalent to
a transition between the minima  of the potential well.
Introducing a shift of the initial moment of time $t\rightarrow
t+\theta$   we can obtain the one dimensional parameterization for an
ensemble of the separatrix trajectories and for the distances between the
trajectories   as well. Using the shifted separatrix solutions
\be
\label{eq:shiftsol}
x=\frac{\sqrt{2}}{\cosh(t+\theta)},
\qquad p=-\frac{\tanh(t+\theta)}{\cosh(t+\theta)},
\ee
from (\ref{eq:melnikov})  we obtain:
\be \label{eq:melnikov1}
\Delta(\theta)=\frac{\alpha\pi\sqrt{2}}{\cosh({\pi}/{2})}\cos(\theta).
\ee
From (\ref{eq:melnikov1}) we see that the dimensionless switching time is equal to the $\theta \approx \pi$.
Therefore, taking into account the connection between the real and the dimensionless
time $\theta\longrightarrow \theta \cdot \Omega_{B}$
we conclude   that the real switching time is proportional to the inverse precession rate $\theta \approx \pi/\Omega_{B}
\approx T$. However, note that the  switching happens only close to the separatrix and the system needs
an additional time to reach the separatrix state. The time that the system
needs to reach the separatrix state should be larger than
$\theta\approx\pi/\Omega_{B}$. This can be checked by numerical
calculations as well (see Fig. \ref{Fig:2} and Fig. \ref{Fig:3})

\section{Width of stochastic layer and criteria of chaos}
\begin{figure}[t]
 \centering
  \includegraphics[width=8cm]{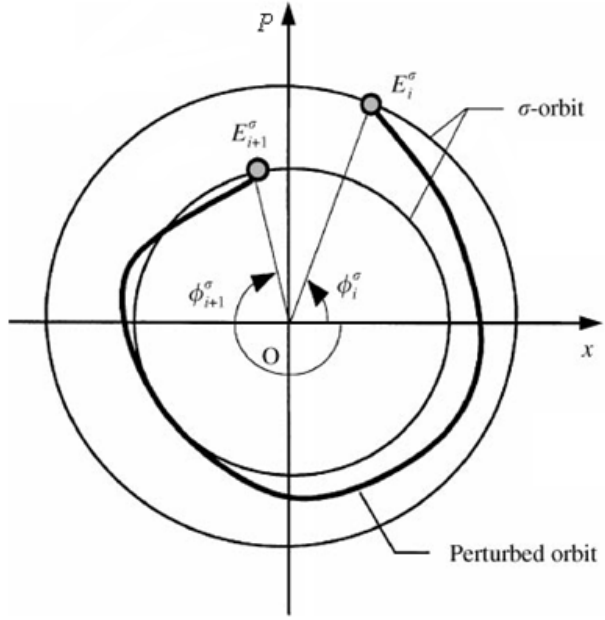}
  \caption{Perturbed trajectories in the vicinity of the separatrix \cite{Gu}. } \label{Fig:4}
\end{figure}

As was mentioned above, the SO interaction should lead to the
formation of a stochastic layer in the vicinity of the separatrix
line \cite{Zaslavsky}. The finite width of the stochastic layer can be
evaluated using the theoretical approach developed in the papers
\cite{Luo,Gu}. Since the separatrix divides the phase space in two parts
$\sigma= (\mbox{in, out})$,  the perturbation $V(t)=\alpha p\cos(\Omega_{B}t)$ leads to the formation of a stochastic layer
on  both sides of the separatrix, see Fig. \ref{Fig:4}. In what follows we refer to the  inner and outward parts of the stochastic layer by the index $(in,out)$, respectively.The inverse precession frequency $1/\Omega_{B}$ defines the time scale of
the dimensionless time and therefore we set $\Omega_{B}=1$.
\newline The change of the energy of the system $H_{0}$ and the angular variable
$\varphi$ during one oscillation period $T_{\sigma}$ forms the
discrete map
$(E_{i};\varphi_{i})\rightarrow(E_{i+1;\varphi_{i+1}})$,
$t_{i}\rightarrow t_{i+1}=t_{i}+T_{\sigma}$,    $\sigma=({\rm in,out})$.
\be \label{eq:map}E_{i+1}=E_{i}+\Delta
H^{\sigma}(\varphi_{i});\quad\varphi_{i+1}=\varphi_{i}+\Delta\varphi^{\sigma}(E_{i+1}),
 \ee
  where $E_{i}=H(x(t_{i}),p(t_{i}))$,
 $\varphi_{i}=\varphi(x(t_{i}),p(t_{i}))$ and $\Delta
 H^{\sigma(\varphi_{i})}$, $\Delta \varphi ^{\sigma}(E_{i+1})$ are the increment of the energy and the phase during one period that
  may occur due to the applied perturbation. The change of the phase and the energy of the
  system $\Delta \varphi^{\sigma}(E_{i+1})$, $\Delta
  H^{\sigma}(\varphi_{i})$ can be evaluated via the following equations
  \be\label{eq:delrafideltaH}
  \Delta \varphi ^{\sigma}\approx
  \Omega_{B}T_{\sigma}(E_{i+1}),\quad \Delta H^{\sigma}(\varphi_{i})\approx
  \int\limits_{t_{i}}^{t_{i}+T_{\sigma}(E_{i})}\{H_{0},V \} dt .
  \ee
  Here $\{H_{0},V \}$ is the Poisson bracket.
   Due to the nonlinearity, the frequency of the orbital motion depends on the energy of the system with
  $\omega_{\sigma}=2 \pi / T_{\sigma}(E)$.
  Therefore, the resonance condition with the external perturbation in case of nonlinear system has the
  form $m_{\sigma}\omega_{\sigma}=n_{\sigma}\Omega_{B}$.
  Here $n_{\sigma}$, $m_{\sigma}$ are integer numbers.The fixed points for the particular
  resonance $(m_{\sigma}:n_{\sigma})$ are defined via the following
  equations:
\be \label{eq:fixed}
E_{i+1}=E_{i}=E_{\sigma}^{m_{\sigma}:n_{\sigma}}, \ee \be
\Delta\varphi^{\sigma}(E_{\sigma}^{(m_{\sigma}:n_{\sigma})})
\approx \frac{2\pi m_{\sigma}}{n_{\sigma}}, \qquad \Delta
H^{\sigma}(\varphi_{\sigma}^{(m_{\sigma}:n_{\sigma})})=0.
\ee
From (\ref{eq:fixed}) one can define resonant values of the phase
$\varphi_{\sigma}^{(m_{\sigma}:n_{\sigma})}$ and energy
$E_{\sigma}^{(m_{\sigma}:n_{\sigma})}$. Energy of the system
itself can be presented as a sum of the resonant part
$E_{\sigma}^{(m_{\sigma}:n_{\sigma})}$ and deviation $\Delta
E_{i}$
\be \label{eq:deviation}
E_{i}=E_{\sigma}^{(m_{\sigma}:n_{\sigma})}+\Delta E_{i}.
\ee
Using
(\ref{eq:deviation}) we can rewrite map (\ref{eq:map}) in the
form
\begin{eqnarray}\label{eq:map1}
&&\Delta E_{i+1}=\Delta E_{i}+\Delta H^{\sigma}(\varphi_{i}),\\
&&
\varphi_{i+1}=\varphi_{i}+\Delta\varphi^{\sigma}(E_{i+1})=\varphi_{i}+\frac{\partial\Delta\varphi(E_{i+1})}{\partial
E_{i+1}}\Bigg|_{E_{i+1}=E_{\sigma}^{(m_{\sigma}:n_{\sigma})}}\Delta
E_{i+1}.\nonumber
\end{eqnarray}
With the notations
\begin{eqnarray}
&& G_{\sigma}^{(m_{\sigma}:n_{\sigma})}=\frac{\partial \Delta
\varphi (E_{i+1})}{\partial E_{i+1}}\Big
|_{E_{i+1}=E_{\sigma}^{(m_{\sigma}:n_{\sigma})}},  \\
&& I_{i}=G_{\sigma}^{(m_{\sigma}:n_{\sigma})}\cdot \Delta E_{i},
\nonumber
\end{eqnarray}
Eq. (\ref{eq:map1}) can be rewritten in the following form
\begin{eqnarray}\label{eq:map2}
&&I_{i+1}=I_{i}+G_{\sigma}^{(m_{\sigma}:n_{\sigma})}\cdot \Delta
H^{\sigma} (\varphi_{i}),\\
&& \varphi_{i+1}=\varphi_{i}+I_{i+1}.\nonumber
\end{eqnarray}
Using (\ref{eq:motion}), (\ref{eq:shiftsol}) and
(\ref{eq:delrafideltaH}) one can derive the explicit expression for
the change of the energy of the system $H_{0}$, during one
oscillation period $T_{\sigma}$:
\begin{eqnarray}\label{eq:change
energy}
&&\Delta H^{\sigma}(\varphi_{i})\approx\int\limits_{t_{i}}^{t_{i}+T_{\sigma}(E_{i})}\big\{H_{0},V\big\}dt=\\
&&=\alpha\left(\int\limits_{-\infty}^{+\infty}dt\frac{2\sqrt{2}\cos\Omega_{B}
t}{\cosh^3[\Omega_{B}(t-t_{i})]}
-\sqrt{2}\int\limits_{\infty}^{+\infty}dt
\frac{2\sqrt{2}\cos\Omega_{B} t}{\cosh[\Omega_{B}(t-t_{i})]}\right)
=\alpha\sqrt{2}\frac{\pi}{\cosh\left({\pi}/{2}\right)}\cos\varphi_{i},\nonumber\\
&&\sigma=({\rm in, out}),\quad\varphi_{i}=t_{i}.\nonumber
\end{eqnarray}
In (\ref{eq:change energy}), due to the time localization  of the
profile of the solution  (\ref{eq:shiftsol}), we extended the limits of
integration to infinity.  For the evaluation of the phase increment
$\Delta\varphi^{\sigma}(E_{i+1})$ that occurs during one period,
again we note that the separatrix divides the phase space on two parts
$\sigma=(\mbox{in,out})$. For inner area  $\sigma={\rm in}$ the period of
the oscillation, for the solution (\ref{eq:solition}) reads
$T_{0}={2\sqrt{2}}K(k)/{x_{0}}$. Therefore, for the phase increment
during one period of the oscillation we have
$\Delta\varphi(E_{\alpha})=T_{0}={2\sqrt{2}}K(k)/{x_{0}}$.  Taking
into account the logarithmic divergence of the elliptic integral
$K(k\approx\ln\left({16}/({1-k^2})\right)/2$ and the
relations $x_{0}=\sqrt{2}$, $1-k^2=-E_{in}=|E_{in}|$, the
phase increment that occurs during one oscillation period, for the
trajectory in the vicinity of the separatrix reads
$\Delta\varphi(E_{in})=T_{in}=\ln\left({16}/{|E_{in}|}\right)$.
Taking into account this fact, for the explicit form of the map
(\ref{eq:map}) we deduce
\begin{eqnarray}\label{eq:mapexplicit}
&&\varphi_{i+1}=\varphi_{i}+\ln\left(\frac{16}{|E_{i+1}|}\right),\\
&&E_{i+1}=E_{i}+\alpha
\sqrt{2}\frac{\pi}{\cosh({\pi}/{2})}\cos\varphi_{i}.
\end{eqnarray}
In particular, for the fixed points of the first
resonance   $(m_{\sigma}:1_{\sigma})$ from (\ref{eq:mapexplicit})
we deduce:
\be\label{eq:firstreson}
E_{i+1}=E_{i}=E_{in}^{m:1},~~~~~\varphi_{i+1}=\varphi_{i}+2\pi
m=\varphi_{in}^{m:1}+2\pi m.
\ee From
(\ref{eq:firstreson}) we obtain
\be \label{eq:rescond}
\alpha\sqrt{2}\frac{\pi}{\cosh\left({\pi}/{2}\right)}\cos\varphi_{in}^{m}=0, \qquad
\ln\left(\frac{16}{|E_{in}^{m}|}\right)=2m\pi
.\ee
Taking into account (\ref{eq:rescond}) we find
 \be \label{eq:efen}
 \varphi_{\alpha}^{m}=\frac{\pi}{2}+k\pi, k\in N,\quad |E_{\alpha}^{m}|=16e^{-2m\pi}.
 \ee
  Using (\ref{eq:efen}) we can rewrite the map (\ref{eq:mapexplicit}) in the vicinity of the fixed points in the following form
 \begin{eqnarray}\label{eq:universal}
 && I_{i+1}=I_{i}-K_{m} \sin\theta_{i}, \\
 && \theta_{i+1}=\theta_{i}+I_{i+1}.\nonumber
 \end{eqnarray}
 Here
 \be \label{eq:K}
 K_{m}=\frac{\alpha\sqrt{2}}{|E_{in}^{m}|}\frac{\pi}{\cosh\left({\pi}/{2}\right)},
 \ee
 is the coefficient of the stochasticity and for convenience we used  the new variable with the shifted
 phase  $\theta_{i}=\varphi_{i}+{\pi}/{2}$.
 The continuous limit of the map (\ref{eq:universal}) reads
 \begin{eqnarray}\label{eq:contlimit}&&\frac{dI}{di}=-K_{m}\sin\theta ,\\
&& \frac{d \theta}{d i}=I, \nonumber
 \end{eqnarray}
 where the index $i$ plays the role of time.
 The differential equations (\ref{eq:contlimit}) can be derived straightforwardly   from the effective
 Hamiltonian
 \be\label{eq:effhamilt}
 H_{\rm eff}=\frac{1}{2}I^2+K_{m}\sin\theta
 .\ee
   \begin{figure}[t]
 \centering
  \includegraphics[width=14cm]{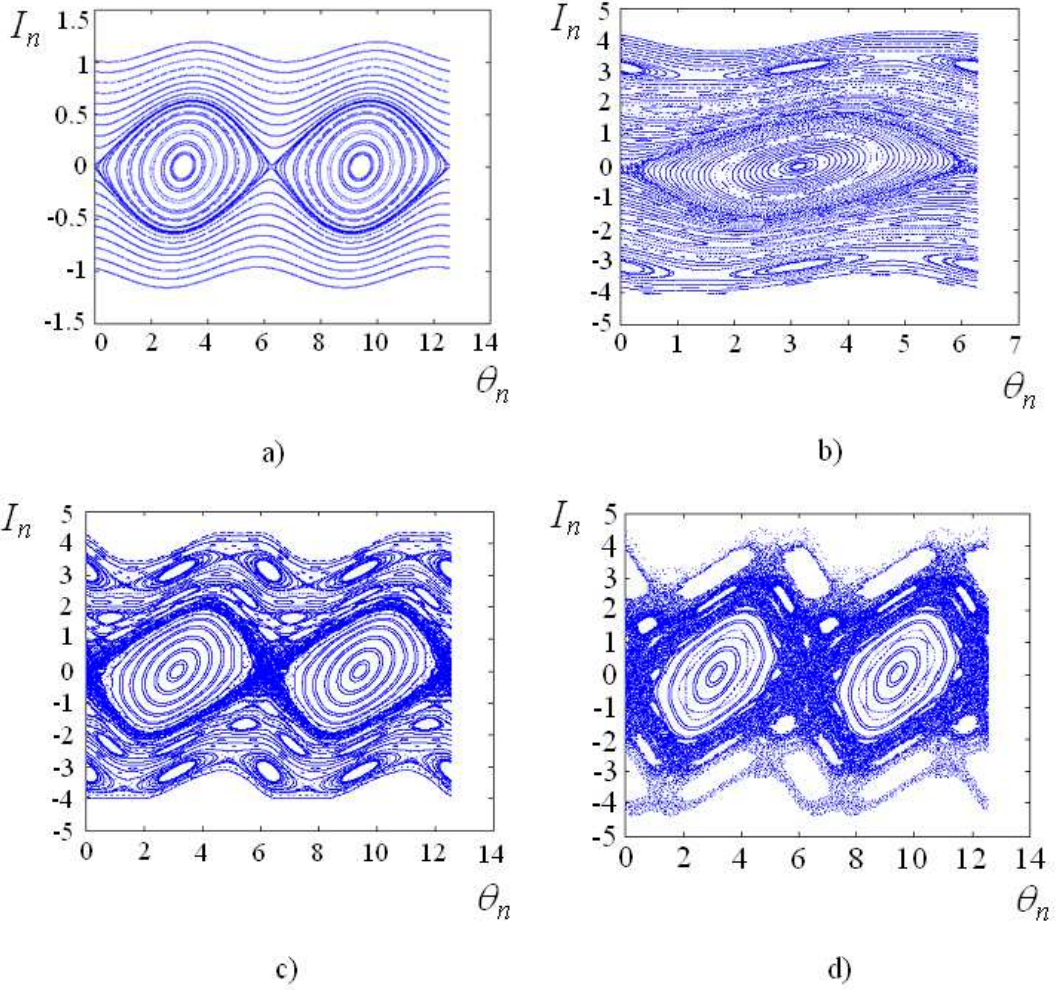}
  \caption{The phase portrait of the system given by eq. (\ref{eq:universal}).
  About $200$ trajectories are generated for the different values of the stochastic parameter $K$:
a) $K=0.1$, b) $K=0.7$, c) $K=1$, d) $K=1.2$.} \label{Fig:5}
\end{figure}
On  the other hand, the effective Hamiltonian responsible for the map (26) reads
\be
H'_{\rm eff}=\frac{1}{2}I^{2}-K_{m}\cos\theta\sum\limits_{\nu=-\infty}^{\infty}\delta(t- \nu).
\ee
The separatrix energy for the Hamiltonian (\ref{eq:effhamilt}) is
equal to the coefficient of the stochasticity for the map
(\ref{eq:universal}) $E_{s}=K_{m}$ \cite{Zaslavsky}. In addition, we note that the equivalence  of our system (26) to the effective Hamiltonian (30) is very helpful. Since for the evolution of the width of the stochastic layer for our system we simply can adopt the well-known results for the system \cite{Sagdeev}  (30).
In our derivations we have considered the trajectory that lies in the
vicinity of the separatrix of the initial system (\ref{eq:motion})
(see Fig.\ref{Fig:4}). Therefore, in our derivation
$E_{in}^{m}$ is supposed to be small however different from
zero. The reason is that exactly on the separatrix $E_{in}^{m}=0$
of the initial system (\ref{eq:motion}) a motion is impossible since the
period of the oscillation diverges logarithmically
$T_{in}=\ln\left({16}/{|E_{in}^{m}|}\right)\rightarrow\infty$, $E_{in}^{m}\rightarrow0$.
Consequently, the coefficient of the stochasticity (\ref{eq:K}) is large
but finite. Using (\ref{eq:efen}) we can rewrite the coefficient of the
stochasticity in the following form
\be \label{eq:Kstoc}
K_{m}=\frac{\alpha\sqrt{2}}{16}e^{2m\pi}\frac{\pi}{\cosh\left({\pi}/{2}\right)}.
\ee
From Eq. (\ref{eq:Kstoc}) we can easily infer  the connection between the amplitude of the SO coupling $\alpha$  and the emergence of
chaos.  Chaos in the system appears if  $K_{m}>1$.

 Taking into account (26) and (\ref{eq:Kstoc}) we obtain
 \be\label{eq:alfa}
 \alpha>\sqrt{\frac{4U_{0}}{m_{e}}}\frac{|E_{in}^{m}|}{\pi\sqrt{2}}\cosh\left({\pi}/{2}\right).
 \ee
The width of the stochastic layer reads \cite{Sagdeev}
 \be\label{eq:widthlayer}
 \delta
 E_{m}=2U_{0}(4\pi)^{4}\sqrt{K_{m}}\exp\left(-\frac{\pi^2}{\sqrt{K_{m}}}\right).
 \ee

In (\ref{eq:alfa}) and (\ref{eq:widthlayer}) we have used the
 definitions $E\rightarrow E/4U_{0}$,  $\alpha\rightarrow\alpha\sqrt{{m_{e}}/{4U_{0}}}$ and from the
 dimensionless SO constant and the dimensionless energy we switched back to the
  real SO constant and the energy. The value of the width of the stochastic layer is a very important quantity.
  Since $\delta E_{m}$ defines the width of the energy interval where chaos appears $E>-\delta E_{m}<$. Note that the width of the stochastic layer $\delta E_{m}$ depends on the two real physical parameters only: The barrier height $U_{0}$ and the SO coupling constant $\alpha$.
  Therefore, the  energy  when chaos appears in the system
  can be estimated easily and verified by the experiment. The width of the stochastic layer is proportional to the SO constant. Consequently, the energy window of the chaotic dynamics $-\delta E_{m}<E<$ is proportional to the SO constant as well.  From (\ref{eq:alfa}) we see that the chaos criteria $K_{m}>1$, connect to  several real physical parameters such as the SO coupling constant $\alpha$, the height of the potential barrier  $U_{0}$, energy of the system, and the effective mass of the electron $m_{e}$. Therefore, the  effect of chaos should be easily observable in the experiment by tuning these parameters.
Close to the separatrix when the energy of the system tends to zero $E_{in}^{m}\rightarrow 0$  the chaos
criteria $K_{m}>1$  holds even for the very weak SO interaction which confirms the   sensitivity of
  the system near  the separatrix. The simple relations (\ref{eq:K})-(\ref{eq:widthlayer}) naturally define  a particular class
of the materials where the transition from the regular to the irregular orbital dynamics of the
  electron can be observed easily on the experiment. A phase portrait for the orbital motion of the electron, for the
different values of the coefficient of the stochasticity is shown  in
   Fig.\ref{Fig:5}.  As we see from Fig. 5 with the increase of the coefficient $K=K_{m}$, the system undergoes a
transition from the regular motion Fig.\ref{Fig:5}a) to the irregular one displayed in Fig.\ref{Fig:5}b).

\section{Conclusions.}
In the present project we considered a simple model relevant for
spintronics. The  electron spin is coupled to the orbital motion via a
SO coupling term and is confined in a double quantum dot. For the
control of the spin dynamics, we considered a driving protocol based
on an  external magnetic field applied along $z$ axis. We have
shown that if the amplitude of the applied field is strong enough
$B>{2\alpha}\sqrt{2mU_{0}}/{\mu_{B}g}$, the spin dynamics is
periodical in time. In particular, the spin rotates around
$z$ axis with a frequency $\Omega_{B}={\mu_{B}gB}$
and for the $x$ projection of the spin we have
$\sigma_{x}(t)=\sigma_{x}^{(0)}\cos(\Omega_{B}t)$. Due to this the
orbital dynamics can be reduced to an  effective,  time dependent,
one dimensional Hamiltonian model (\ref{eq:ham1}). Different types of the
dynamics can be
realized. If  the energy of the system is negative
$-{1}/{4}<E<0$, the electron is localized in the right or in the
left well and performs  oscillations bound by the
 potential barriers. However, via  the spin orbit coupling channel, energy can
be transferred from the spin system to the orbital motion.  This
naturally heat ups the orbital motion of the electron and   due to the
presence of a  separatrix line in the phase space of the system
(\ref{eq:motion}),  the electron may change from one potential
well to the other.
We call this effect a spin orbit coupling induced "tunneling" (even though this process is not a tunneling in the
quantum mechanical sense)
to stress the fact that the electron can perform inter-well transitions.
 We derived simple
analytical equations (\ref{eq:K})-(\ref{eq:widthlayer}) for the appearance of chaos
 and showed how this regime can be reached via tuning    real physical parameters
such as  SO coupling constant $\alpha$, the height of the potential
barrier $U_{0}$ , the  energy of the system, and the effective mass of the
electron $m_{e}$ . Therefore, the  effect of chaos should be easily
observable on the experiment. We also proved that close to the
separatrix when the energy of the system tends to zero
$E_{in}^{m}\rightarrow0$ the chaos criteria $K_{m}>1$ holds even
for a very weak SO interaction, confirming by this the extreme
sensitivity of the system near to the separatrix. In addition we
derived expression for the width of the stochastic layer
(\ref{eq:widthlayer}). We proved that the width of the separatrix is
proportional to the SO constant and inversely proportional to the
energy of the system.
 \newline \textbf{Acknowledgments} We thank  E.Ya. Sherman for useful discussions. The financial support
by the Deutsche Forschungsgemeinschaft (DFG) through SFB 762,
contract BE 2161/5-1, Grant No. KO-2235/3, and STCU Grant No. 5053
is gratefully acknowledged.

\end{document}